\documentclass{emulateapj}

\newcommand{\Msun}{\ensuremath{M_{\odot}}}
\newcommand{\nthp}{N$_2$H$^+$}

\usepackage[backref,breaklinks,colorlinks,citecolor=blue]{hyperref}%backref
\usepackage[all]{hypcap}

\slugcomment{Submitted to ApJ}

\shorttitle{Virial Stars from Sub-Virial Cores}
\shortauthors{Foster, J.B. et al.}

\begin{document}

\title{IN-SYNC II: Virial Stars from Sub-Virial Cores -- The Velocity Dispersion of Embedded Pre-Main-Sequence Stars in NGC 1333}

\author{
Jonathan B. Foster\altaffilmark{1}, 
  \and Michiel Cottaar\altaffilmark{2},
  \and Kevin R. Covey\altaffilmark{3,4}, 
  \and H\'{e}ctor G. Arce\altaffilmark{5},
  \and Michael R. Meyer\altaffilmark{2},
  \and David L. Nidever\altaffilmark{6}, 
  \and Keivan G. Stassun\altaffilmark{7,8}, 
  \and Jonathan C. Tan\altaffilmark{9}, 
  \and S. Drew Chojnowski\altaffilmark{10}, 
  \and Nicola da Rio\altaffilmark{9}, 
  \and Kevin M. Flaherty\altaffilmark{11}, 
  \and Luisa Rebull\altaffilmark{12},
  \and Peter M. Frinchaboy\altaffilmark{13}, 
  \and Steven R. Majewski\altaffilmark{10},
   \and Michael Skrutskie\altaffilmark{10}, 
   \and John C. Wilson\altaffilmark{10},
   \and Gail Zasowski\altaffilmark{14},
}
  \altaffiltext{1}{Yale Center for Astronomy and Astrophysics, Yale University, New Haven, CT 06520, USA; \email{jonathan.b.foster@yale.edu}
}
  \altaffiltext{2}{Institute for Astronomy, ETH Zurich, Wolfgang-Pauli-Strasse 27, 8093 Zurich, Switzerland}
  \altaffiltext{3}{Lowell Observatory, Flagstaff, AZ 86001, USA}
  \altaffiltext{4}{Current Address: Dept. of Physics \& Astronomy, Western Washington Univ., 516 High Street, Bellingham WA 98225, USA}
  \altaffiltext{5}{Department of Astronomy, Yale University, P.O. Box 208101, New Haven, CT 06520, USA}
  \altaffiltext{6}{Department of Astronomy, University of Michigan, Ann Arbor, MI 48109, USA}
  \altaffiltext{7}{Department of Physics \& Astronomy, Vanderbilt University, VU Station B 1807, Nashville, TN, USA}
   \altaffiltext{8}{Physics Department, Fisk University, Nashville, TN 37208, USA}
  \altaffiltext{9}{Department of Astronomy, University of Florida, Gainesville, FL 32611, USA}
  \altaffiltext{10}{Department of Astronomy, University of Virginia, Charlottesville, VA 22904, USA}
  \altaffiltext{11}{Astronomy Department, Wesleyan University, Middletown, CT, 06459, USA}
   \altaffiltext{12}{Spitzer Science Center/Caltech, 1200 E. California Blvd., Pasadena, CA 91125, USA}
  \altaffiltext{13}{Department of Physics \& Astronomy, Texas Christian University, Fort Worth, TX 76129, USA}
  \altaffiltext{14}{Department of Physics and Astronomy, Johns Hopkins University, Baltimore, MD 21218, USA}

\begin{abstract}
The initial velocity dispersion of newborn stars is a major unconstrained aspect of star formation theory. Using near-infrared spectra obtained with the APOGEE spectrograph, we show that the velocity dispersion of young (1-2 Myr) stars in NGC 1333 is 0.92$\pm$0.12~km~s$^{-1}$ after correcting for measurement uncertainties and the effect of binaries. This velocity dispersion is consistent with the virial velocity of the region and the diffuse gas velocity dispersion, but significantly larger than the velocity dispersion of the dense, star-forming cores, which have a sub-virial velocity dispersion of 0.5~km$^{-1}$. Since the NGC 1333 cluster is dynamically young and deeply embedded, this measurement provides a strong constraint on the initial velocity dispersion of newly-formed stars. We propose that the difference in velocity dispersion between stars and dense cores may be due to the influence of a 70$\mu$G magnetic field acting on the dense cores, or be the signature of a cluster with initial sub-structure undergoing global collapse.

\end{abstract}


\section{Introduction}

The initial velocity of a newborn star is one of a few fundamental stellar properties set by the star-formation process; as such, it can provide powerful constraints on theories and simulations of star-formation. The 3D velocities of young stars can be assessed by measurements of radial velocities or proper motions, either of which allows an estimate of the stellar velocity dispersion as well as, in principle, global motions such as expansion/contraction and rotation. The radial velocities of young stars are particularly useful as these velocities can be directly compared to the (radial) velocity of the molecular gas in which they are embedded. 

Previous work has shown that dense gas cores (both starless and hosting protostars) have a lower velocity dispersion than the diffuse gas in which they are embedded, and out of which these dense cores presumably formed \citep{Walsh:2004, Andre:2007, Kirk:2007, Rosolowsky:2008, Kirk:2010}. In regions which form predominantly low-mass stars, the dense cores have a typical one-dimensional velocity dispersion of 0.4 to 0.5 km~s$^{-1}$.

Studies of the radial velocity of stars in optically-revealed star-forming regions also find that the velocity dispersion of the stars is lower than the diffuse gas in the same region. For instance, in the Orion Nebula Cluster (ONC) and NGC 2264, the one-dimensional velocity dispersion of stars is 3-4 km s$^{-1}$ \citep{Furesz:2006, Furesz:2008, Tobin:2009}. 

There are important distinctions between the young stellar velocities measured in the ONC and NGC 2264 and the dense gas core velocities in nearby low-mass regions. First, the ONC and NGC 2264 are significantly more massive and have a larger virial velocity dispersion. Second, the dynamical time in these regions is short, so although the stars are only 1-3 Myr old, their velocities may have evolved over 4-12 dynamical times \citep{Tan:2006}. Finally, the fact that one is able to observe these stars in the optical suggests that they have significantly dispersed their natal gas, which can profoundly affect the stellar dynamics \citep[e.g.][]{Moeckel:2010}. For these reasons the measurements in optically-revealed clusters do not directly reveal the initial velocity dispersion of new stars.

\citet{Covey:2006} used near-infrared spectra of Class I and flat-spectrum objects within a number of nearby star-forming regions to show that these young stars have velocity dispersions similar to, or slightly larger than, the gas in which they are embedded. The velocity dispersions measured by \citet{Covey:2006} were comparable to the 1.5~km~s$^{-1}$ radial velocity precision of their observations, however, and thus provide only an upper limit on the true velocity dispersion of these protostars.

In a companion paper to this one, \citet{Cottaar:2014c} show that the 2-6 Myr old, optically-revealed cluster IC 348 has a velocity dispersion of 0.6 - 0.7~km~s$^{-1}$, which is slightly super-virial.

Simulations of the dynamical evolution of young clusters \citep[e.g.][]{Proszkow:2009, Moeckel:2012, Parker:2012, Girichidis:2012, Kruijssen:2012} show that the cluster's initial conditions can be quickly erased by dynamical evolution. In \citet{Proszkow:2009}, for example, two simulations are compared, starting the stars with either a sub-virial or a virial velocity distribution. The sub-virial distribution collapses in size and increases its velocity dispersion within 1.5 Myr. Stars this young are normally still embedded, so near-infrared spectroscopy is necessary to measure their stellar properties and velocities. 

Obtaining these high resolution near-infrared spectra is the goal of the INfrared Spectra of Young Nebulous Clusters (IN-SYNC) project \citep{Cottaar:2014b}, which is measuring stellar velocities with a precision of $\sim$0.3 km~s$^{-1}$ for IC 348 \citep{Cottaar:2014c} and NGC 1333 in Perseus, as well as the more massive regions NGC 2264 and Orion A \citep{DaRio:inprep}. IN-SYNC is an ancillary science program of the Apache Point Observatory Galactic Evolution Experiment \citep[APOGEE;][]{Zasowski:2013, Majewski:inprep}, part of the third Sloan Digital Sky Survey \citep[SDSS-III;][]{Gunn:2006, Eisenstein:2011}.

The cluster NGC 1333 in Perseus presents us with an excellent location to compare the velocity dispersion of dense cores and young stars within a single region. NGC 1333 is young enough to contain both dense cores and young stars \citep[$\sim$ 1 Myr;][and references therein]{Arnold:2012}, near enough to resolve the dense gas cores \citep[$\sim$~250 pc;][]{Hirota:2008, Bell:2013, Plunkett:2013}, yet not so massive that the embedded pre-main sequence population is rendered inaccessible by dust extinction even in the near-infrared. NGC 1333 therefore affords us the opportunity to build a full picture of the velocities of the diffuse gas, the dense cores, and the young embedded stars within a single cluster.

\section{Observations}
\subsection{Stellar Data}

\begin{figure}
\includegraphics[width=0.49\textwidth]{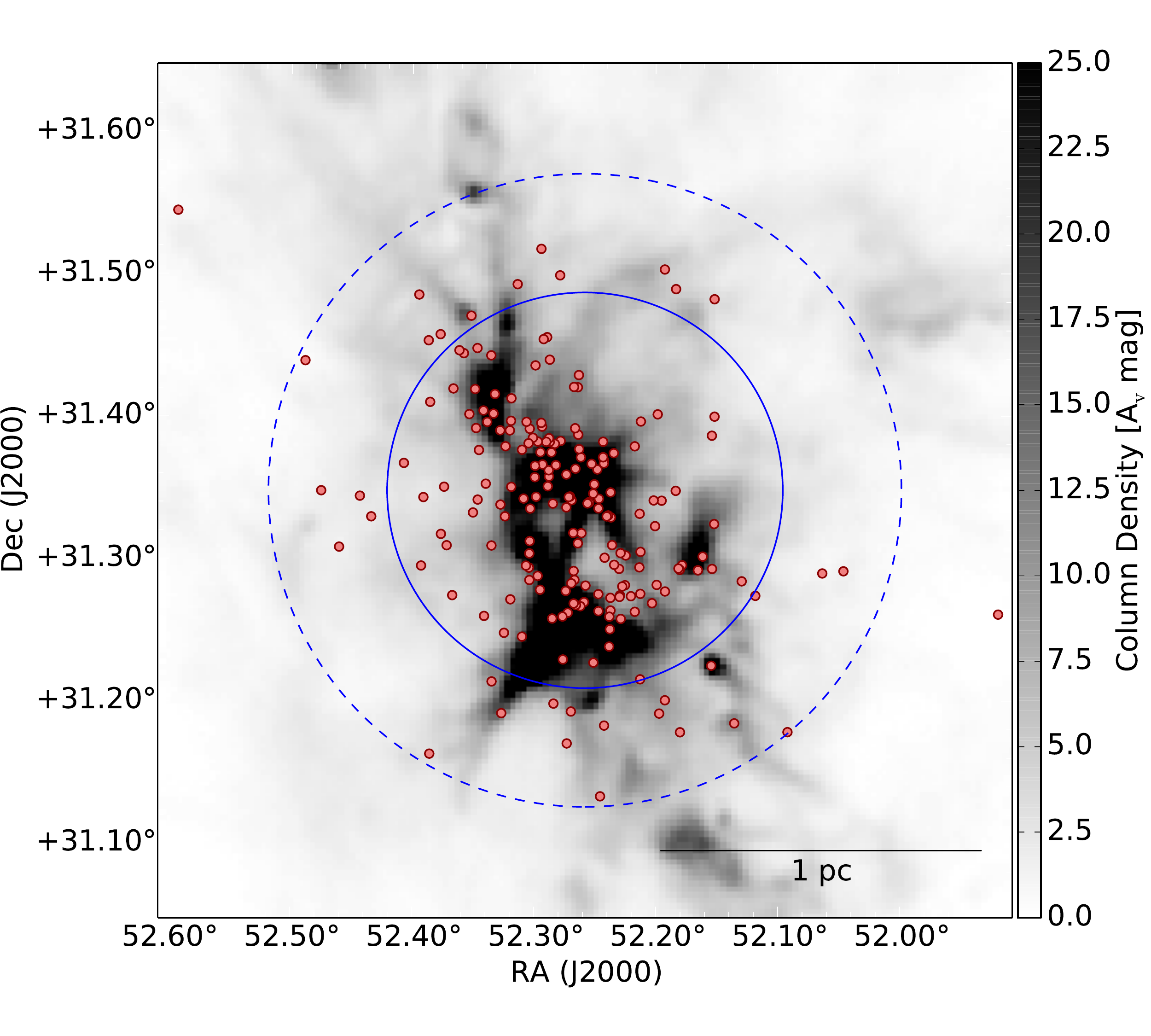}
\caption{Column density map of NGC 1333 derived from \emph{Herschel} data (grayscale). Red circles show the full IN-SYNC catalog. Fiducial radii are shown corresponding to the boundary of the dense gas (solid blue line) and the majority of stellar population (dashed blue line). }
\label{fig:coldensity}
\end{figure}

\begin{figure}
\includegraphics[width=0.44\textwidth]{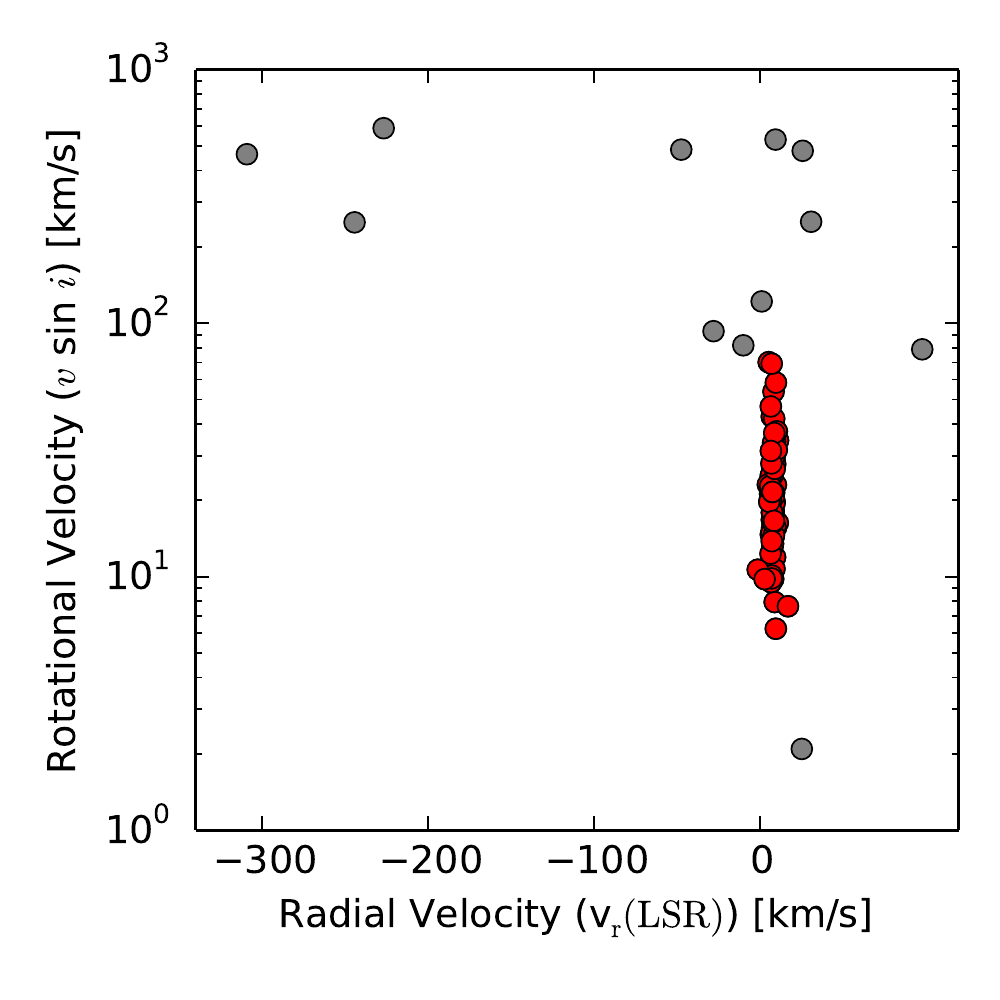}
\caption{Radial velocity versus rotational velocity for the stars in NGC 1333 following the SNR and reduced $\chi^2$ cuts described in \autoref{sec:stellarprops}. The abrupt increase in radial velocity dispersion above 75~km~s $^{-1}$ corresponds to early-type stars with broad intrinsic line widths; the single star with $v$ sin $i$ $<$ 5~km~s $^{-1}$ is likely a field star. We keep only the intermediate stars (shown in red) for further analysis. }
\label{fig:velcut}
\end{figure}

\begin{figure*}
\includegraphics[width=0.99\textwidth]{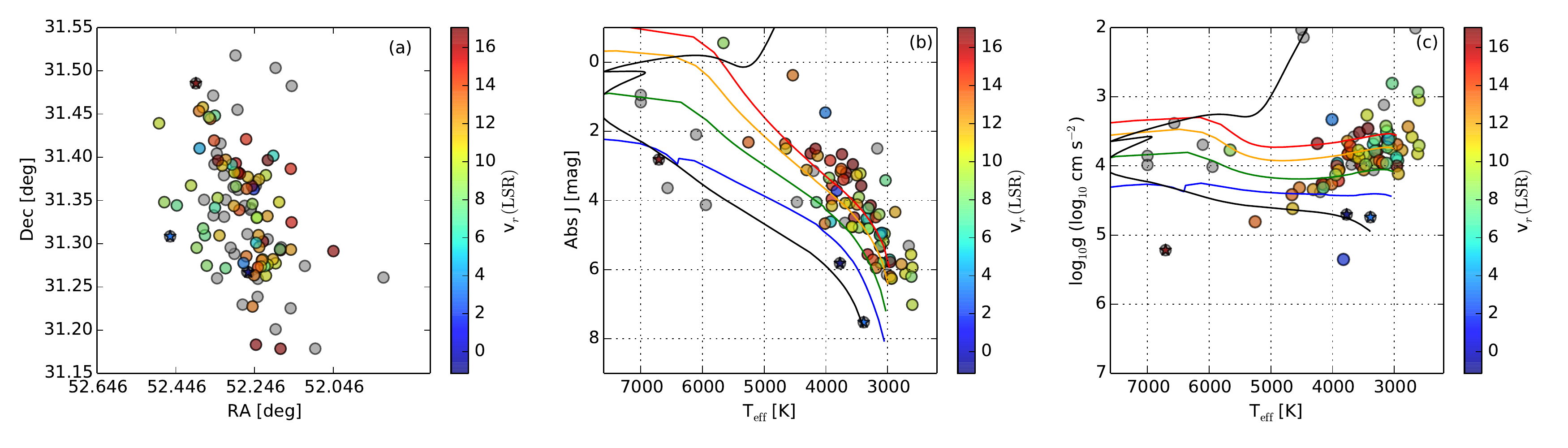}
\caption{Sample of pre-main-sequence stars in NGC 1333 considered in this study. (a) The spatial distribution of stars, both included (color-coded by radial velocity) and not included (gray) in the sample based on the quality criteria given in the text. (b and c) Stars included and not included, with pre-main-sequence isochrones in color at 1 (red), 2 (orange), 6 (green) and 10 (blue) Myr and a 1 Gyr main-sequence isochrone (black); all isochrones from \citet{Dotter:2008}. The diagram of absolute $J$ magnitude versus effective temperature shows a relatively tight clustering around 1-2 Myr, and is used to exclude non-cluster main-sequence stars (denoted with a star symbol) lying below the 10 Myr isochrone.}
\label{fig:hrd}
\end{figure*}

Spectra were obtained using the APOGEE multi-object spectrograph \citep{Wilson:2012}. The details of the data reduction and spectral fitting are presented in \citet{Cottaar:2014b}. In brief, a grid of BT-Settl \citep{Allard:2012} model spectra are convolved to the resolution of the reduced IN-SYNC spectra. The main parameters that are allowed to vary in fitting to the models to the observed spectra are the effective temperature ($T_{\mathrm{eff}}$), the veiling ($r_H$), the rotational velocity ($v \sin{i}$), the surface gravity ($\log(g)$) and the radial velocity ($v_r$) of the star. Initial parameter uncertainties are estimated from the Markov Chain Monte Carlo (MCMC) fitting process, although these uncertainties are then inflated to match the actual epoch-to-epoch variability seen in the parameters.

IN-SYNC targets in NGC 1333 were chosen from the Cores to Disks (c2d) Spitzer survey of Perseus \citep{Jorgensen:2006, Rebull:2007}, supplemented with focused surveys of NGC 1333 \citep{Getman:2002, Gutermuth:2008, Winston:2009, Winston:2010} and candidate members selected based on their mid-IR variability from a preliminary analysis of the light curves obtained in NGC 1333 by the Spitzer YSOVAR program \citep{Rebull:2014a}.

Stars from all input surveys were given equal weight when assigning targets, and the primary selection criteria was $H$ magnitude. Since the APOGEE spectrograph has a fiber collision limit of 71\arcsec, we used three distinct plates in order to achieve nearly complete coverage of bright stars in the center of the cluster. We prioritized assigning fibers to stars with 7.5 mag $<$ $H$ $<$13.5 mag; this was limited at the faint end by signal-to-noise considerations and at the bright end by the potential for flux bleeding. Sources with $H$$<$ 12.5 magnitudes were considered highest priority targets, and then fainter sources were used to fill up a plate. A faint source that was selected for one plate was prioritized on subsequent plates so as to build up the signal-to-noise of faint stars.

This fiber assignment scheme ensured that only five 5 NGC 1333 candidate members with $H$$<$13.5 mag were not assigned a fiber on any of the NGC 1333 fiber plug plates. Otherwise, 141 likely NGC 1333 members with $H$$<$13.5 mag were assigned fibers on at least one NGC 1333 plate; of these, 107 were assigned fibers on two or more plates, with 79, 61, and 58 members assigned at least three, four or five fibers across both observing seasons. The observations are therefore close to complete for the bright stars in NGC 1333, but are significantly incomplete at fainter magnitudes, and biased against faint stars in the densest portions of the cluster.

These magnitude limits do not correspond to simple limits on stellar mass, as the intrinsic $H$-band luminosity of a star in NGC 1333 may significantly effected by extinction, either local (the envelope around a very young star) or global (from substructure within the dust and gas in the cluster as a whole). The former constraint limits the IN-SYNC sample to relatively older stars. Thus, the vast majority of IN-SYNC stars with a \citet{Gutermuth:2008} classification from Spitzer are Class II, rather than Class I stars, simply because local extinction around Class I stars renders them very faint in $H$. For the typical 1 Myr old star in the sample, the $H$-band magnitude limit of 13.5 corresponds to roughly 0.1 \Msun (see \autoref{sec:stellarprops}).

IN-SYNC observations of NGC 1333 include observations of stars in the ``West-End'' of Perseus well outside of NGC 1333. We use a minimal-spanning tree \citep[MST;][]{Gutermuth:2009} on the positions of candidate protostars in the region to define a boundary for the cluster. This method shows a break in the cumulative distribution of span-length at 0.15 degrees, or 0.65 pc. Cutting the MST at this span length produces a cluster boundary that corresponds well to what a by-eye identification of the cluster would provide, and corresponds to all stars within the boundary of the box 51.8$\arcdeg$ $<$ R.A. $<$ 52.5$\arcdeg$ and 31.15$\arcdeg$ $<$ Dec. $<$ 31.6$\arcdeg$. \autoref{table:alltarget} presents the full list of candidate members in NGC 1333 used in this survey, not all of which were observed.

\subsection{Gas Data}
To compare with our stellar velocities, we measure the velocity and velocity dispersion of the local gas with a combination of different tracers. These include the $^{12}$CO (1-0) and $^{13}$CO (1-0) transitions, which trace relatively low volume-density gas; we use the data from the COMPLETE Survey \citep{Ridge:2006a}. In addition, we use maps from higher critical density transitions in the central region of the cluster. These maps were obtained from the JCMT archive and include $^{12}$CO, $^{13}$CO, and C$^{18}$O (3-2) from \citet{Curtis:2011}.  The velocities of dense cores within NGC 1333 are drawn from the \nthp (1-0) observations of continuum sources from \citet{Kirk:2007}. 
  
\subsection{Dust Column Density Map}
\label{sec:dustcolumn}
We have used the publicly available \citep{Andre:2010} \emph{Herschel} data for NGC 1333 to construct a map of the dust column density over NGC 1333. This map was created from fitting the \emph{Herschel} 160 - 500\micron\ data with a single temperature modified black-body where the dust opacities at each wavelength are given by the opacities in \citet{Ossenkopf:1994} for dust grains with thin ice mantles, rather than assuming a simple power-law modification of the blackbody spectrum (i.e., taking a single value of $\beta$). Choosing instead to use $\beta$=2, normalized at 230 GHz \citep[a typical assumption, see][]{Schnee:2010} produces a dust mass that is 10\% greater. For the \citet{Ossenkopf:1994} model the opacity at 500 \micron\, $\kappa_{500}$ is 0.05 cm$^2$ g$^{-1}$; we assume a gas-to-dust ratio 100:1. 

To account for large-scale gradients present in the \emph{Herschel} data, the zero-point of this map was set by matching the column densities obtained around the edges of NGC 1333 with the COMPLETE \citep{Ridge:2006a} extinction map based on 2MASS photometry. The COMPLETE extinction map is lower resolution than our new \emph{Herschel} column density map (2.5\arcmin\ versus 36\arcsec), missing structure evident in the \emph{Herschel} column density map. Additionally, the extinction map is also significantly biased at the position of the cluster by the presence of many embedded red stars. The \emph{Herschel} dust column density map, anchored by the reliable (i.e., non-cluster) portions of the extinction map, therefore provides the best available tracer of the dust (and therefore gas) mass in NGC 1333. For comparison, the total mass of the cluster gas is 20\% lower when estimated from the COMPLETE extinction map rather than the \emph{Herschel}-derived column density map.
 
The column density map derived from \emph{Herschel} is shown in \autoref{fig:coldensity} along with all the stars in the IN-SYNC target catalog; the stars observed by IN-SYNC are a subset of these objects.
 
\subsection{Selecting a Sample of Stars for Analysis}

\label{sec:stellarprops}
\autoref{table:allspectra} shows the best-fit stellar parameters for all APOGEE spectra obtained in NGC 1333. For this analysis we have applied quality criteria to these fits. Specifically, we exclude all spectra with a S/N $<$ 20, and stars for which the reduced $\chi^2$ of the best model fit was $>$ 10, as these spectra lead to unreliable parameter determinations. 

We then trim stars with very low or high rotational velocity ($v$ sin $i$ $<$ 5 km~s$^{-1}$ or  $v$ sin $i$ $>$ 75 km~s$^{-1}$). The former cut removes only one star, 2M03291184+3121557, which is near the edge of the cluster and has a velocity far from the cluster mean; this is likely a contaminating field star. The latter cut may remove genuine cluster members, typically very early type stars dominated by hydrogen lines. The large $v$ sin $i$ for these stars means that they have broad absorption lines; these broad lines mean that the precision of the radial velocity fit is low, and therefore the stars exhibit a much broader spread in radial velocity than stars with more secure fits (see the abrupt increase in velocity spread in \autoref{fig:velcut}). 

These cuts on $v$ sin $i$ are similar to the cuts applied in \citet{Cottaar:2014c} except that we adopt an upper limit of $v$ sin $i$ $=$ 75 km~s$^{-1}$ rather than 150 km~s$^{-1}$, as this corresponds to the observed $v$ sin $i$ threshold for dramatically increased radial velocity scatter in NGC 1333. The poor radial velocity fits for these hot early-type stars means that we are unable to make any conclusive statements about the velocity dispersion at the high mass/effective temperature end of the distribution. Using the \citet{Dotter:2008} pre-main-sequence isochrones and an age of 1 Myr (see \autoref{fig:hrd}), this cut effectively corresponds to removing all stars with M $\gtrsim$ 3.5 \Msun.

We remove all stars from the sample with a radial velocity uncertainty greater than 1.1 km~s$^{-1}$, as these stars contribute little information about the velocity dispersion. We also remove stars with strong radial velocity variability. We estimate this in the same manner as \citet{Cottaar:2014c}, by calculating the $\chi^2$ statistic for the model of constant radial velocity as
\begin{equation}
\chi^2 = \sum_i{\frac{(v_i - \mu)^2}{\sigma_i^2}},
\end{equation}
where $v_i$ and $\sigma_i$ are the best-fit radial velocity and associated uncertainty in each epoch, and $\mu$ is the uncertainty-weighted mean radial velocity. If the probability of obtaining at least this large a $\chi^2$ is less than 10$^{-4}$, we flag the star as a radial velocity variable and exclude it from the following velocity analysis. 6\% of stars are flagged as having variable radial velocities in this fashion.

Finally, after applying these cuts, there are three stars that lie closer to the main-sequence isochrone than the 1-2 Myr isochrone where most of the target clusters (see \hyperref[fig:hrd]{Figure~\ref*{fig:hrd}b}, where these three stars are shown with star symbols). These three stars (2M03290289+3116010, 2M03293476+3129081, and 2M03295048+3118305) are all candidate members, rather than confirmed young stars, are radial velocity outliers, and two of them lie on the outskirts of the cluster (\hyperref[fig:hrd]{Figure~\ref*{fig:hrd}a}); we therefore conclude that these three stars are most likely contaminating field stars, rather than genuine members of the cluster, and we exclude them from further analysis. This leaves 70 stars for the analysis of the cluster's velocity dispersion.

\autoref{table:allstar} shows the uncertainty-weighted mean parameters for all stars observed by IN-SYNC, along with how many spectra were used in the determination and a flag to indicate stars that were identified as having a variable radial velocity.

For comparison with the gas velocities, we have converted all stellar heliocentric radial velocities into the velocity frame of the gas data. This is the kinematical Local Standard-of-Rest (LSRK), defined as a solar motion of 20 km/s in the direction of $\alpha$(J2000),$\delta$(J2000) = (18:03:50.29, +30:00:16.8). No other velocity correction, such as correction for the gravitational redshift \citep{Pasquini:2011} or convective blueshift \citep{Shporer:2011} has been applied. These corrections, which account for the difference between the velocity of the photosphere and the velocity of the center of mass of the star, are typically on the order of a few hundred m s$^{-1}$ (although the convective blueshift is poorly constrained for young stars). From a comparison with literature results, \citet{Cottaar:2014b} estimates that the systematic uncertainty in the absolute zero point of the IN-SYNC radial velocity system is on the order of 0.5 km s$^{-1}$.

From the \citet{Dotter:2008} isochrones, and the age of the cluster (1-2 Myr) from \hyperref[fig:hrd]{Figure~\ref*{fig:hrd}b}, we can assign a mass to each star based on the effective temperature. The lowest mass calculated for the \citet{Dotter:2008} isochrones is 0.1\Msun\ (corresponding to $T_{\mathrm{eff}}$ = 3000 K), and we do not see many stars with temperatures far below this. The mass function (number of stars per unit mass) increases down to the 0.1\Msun\ limit, suggesting that we are reasonably sampling the masses down to 0.1 \Msun. We are therefore sampling masses between 3.5\Msun\ and down to close to the hydrogen burning limit. We are not sensitive to the velocity distribution of brown dwarfs or early-type stars.

\section{Results}
\subsection{Stellar Velocity Dispersion}
\label{sec:stellar-dispersion}

\begin{figure}
\includegraphics[width=0.47\textwidth]{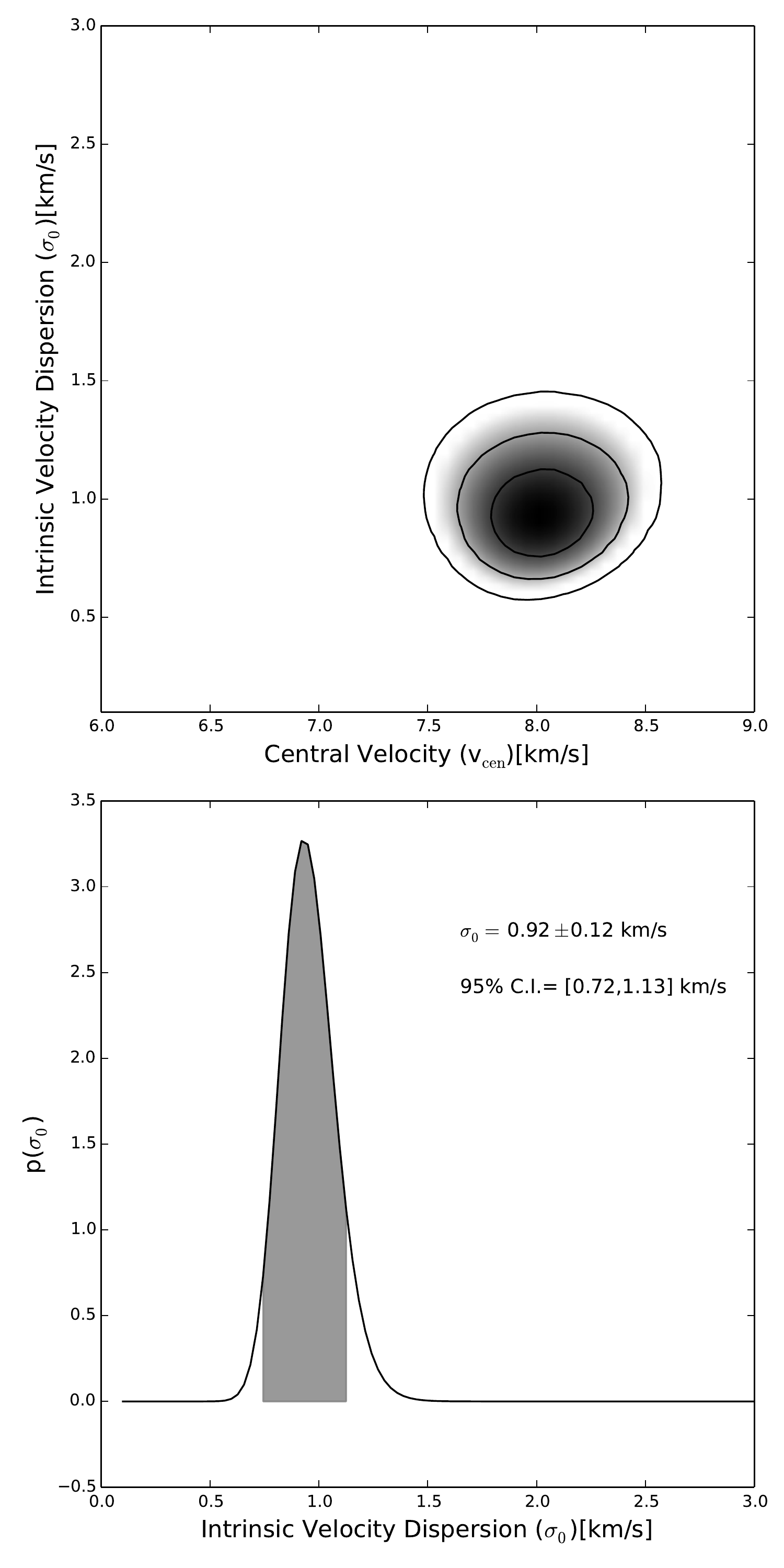}
\caption{The inferred posterior probability distribution from \textsc{velbin} for NGC 1333, marginalized over the fraction of binaries (top) and marginalized over the fraction of binaries and the central velocity (bottom). Contours in the top panel show the 68\%, 95\%, and 99\% credible intervals.}
\label{fig:velbin}
\end{figure}

The determination of the intrinsic velocity dispersion of the young stellar population in NGC 1333 requires two main corrections. First, the radial velocities of the young stars have significant and non-uniform uncertainties. Second, the velocity distribution will be inflated by the presence of binaries, particularly close binaries. Multiple epochs of radial velocity data allow for the identification of some binaries, but not all, particularly since 40\% of the stars in our NGC 1333 sample only have one observation meeting our SNR and goodness-of-fit criteria. 

We take three distinct approaches to infer the intrinsic velocity dispersion: (1) use the \textsc{velbin} package \citep{Cottaar:2014a} to model the velocities at all epochs and the influence of binaries simultaneously, (2) use an outlier-resistant analytic estimate for the velocity dispersion in the case of non-uniform uncertainties, and (3) trim all stars with radial velocity uncertainty greater than 0.5 km~s$^{-1}$ and use outlier-resistant estimates of the width of the distribution. Each of these approaches makes different assumptions, so it is important to check that they produce consistent estimates for the intrinsic velocity dispersion. For comparison, using just the sample standard deviation to estimate the width of the velocity distribution (i.e., ignoring errors and binary contamination) provides an estimate of 1.1$\pm$0.1~km~s$^{-1}$.

We first infer the intrinsic velocity dispersion of NGC 1333 using the \textsc{velbin} package introduced in \citet{Cottaar:2014a}. This package generates a large sample of binary stars, with the mass ratio and orbital properties drawn from literature values. Specifically, we use the log-normal period distribution from \citet{Raghavan:2010}, the nearly-flat mass ratio distribution from \citet{Reggiani:2013}, and the flat eccentricity distribution from \citet{Duchene:2013}. This distribution is then sampled and compared against the observed velocities and velocity errors from the real data set to infer the fraction of sources that are binaries and thus deduce the intrinsic velocity width of NGC 1333's stars after accounting for the influence of binaries. The underlying velocity distribution is assumed to be Gaussian, and so the posterior probability distribution function is inferred for the center of the velocity distribution, the intrinsic velocity dispersion, and the binary fraction. 

\autoref{fig:velbin} shows the posterior probability distribution functions inferred for NGC 1333's velocity distribution, marginalized over the fraction of binaries (which is poorly constrained with a fairly flat posterior distribution between 20 and 70\%) and over both the binary fraction and the central velocity. The latter posterior probability distribution for the intrinsic velocity dispersion, $\sigma_0$, is only slightly asymmetric, with a most likely value of 0.92$\pm$0.12 km s$^{-1}$ and a 95\% credible interval of [0.72, 1.13] km s$^{-1}$. The central velocity, $v_{\mathrm{cen}}$, is 8.02 $\pm$0.31 km s$^{-1}$. For comparison, the centroid velocity of the $^{13}$CO (1-0) gas in this region ranges from 7 to 9 km s$^{-1}$ \citep{Quillen:2005}.

The correction for binarity is relatively small because we have data over a three-year baseline and have already removed 6\% of the stars as radial-velocity variable (and thus likely binaries, see \autoref{sec:stellarprops}). The remaining stars are known not to have a large radial velocity signature over the three years they were observed, and are therefore likely to either (1) not be in short-period binaries, or (2) not have an edge-on inclination that produces a large radial-velocity signature. Since we marginalize over all binary fractions in \autoref{fig:velbin}, this estimate provides a conservative estimate for the intrinsic velocity dispersion, $\sigma_0$. If we fix the binary fraction at extreme values we get the following: for a binary fraction of 80\%, $\sigma_0$=0.89$\pm$0.09 km s$^{-1}$; for a binary fraction of 20\%, $\sigma_0$=0.98$\pm$0.10 km s$^{-1}$.

We also consider the effect of a period cut-off on the log-normal period distribution from \citet{Raghavan:2010}. This has a small influence since the inferred binary fraction increases/decreases in order to match the observational constraints. If the binary fraction is held fixed at 50\%, then there is some sensitivity to adopting period cut-offs. We consider impose a cut-off on the maximum period. A semi-major cut-off of 1 AU provides an estimate of $\sigma_0$ = 0.98$\pm$0.09 km s$^{-1}$, a semi-major cut-off of 10 AU gives $\sigma_0$ = 0.89$\pm$0.10 km s$^{-1}$, and a semi-major cut-off of 100 AU (or more) gives $\sigma_0$ = 0.92$\pm$0.10 km s$^{-1}$. This behavior is because the binaries that can most increase our observed velocity dispersion are binaries with intermediate periods -- short-period binaries produce a large radial velocity signature which is easily ruled out by our multi-epoch observations while long-period binaries produce small radial velocity variations. All these variations are well within our uncertainty given for $\sigma_0$. Our estimate of $\sigma_0$ is therefore relatively robust against changes in the assumed binary population.

The second estimate for the velocity dispersion considers binaries as velocity outliers and infers the parameters of the velocity distribution using robust estimators. For this purpose we use the median and the inter-quartile range to infer the center and width of the distribution. A single velocity for each star is calculated as the weighted mean over all observed epochs, and a single velocity uncertainty is calculated as the median error across the epochs. This approach therefore uses less information than the \textsc{velbin} method.

To account for the non-uniform errors in the epoch-averaged radial velocity measurements, we adopt the outlier-resistant estimator given by \citet{astroMLText}:

\begin{equation}
\sigma_0 = \sqrt{\zeta^2 \sigma_G^2 - e_{50}^2},
\end{equation}
where $\sigma_G$ is the unbiased estimator of $\sigma$ for a Gaussian based on the interquartile range:
\begin{equation}
\sigma_G  = 0.741 (q_{75} - q_{25})
\end{equation}
and
\begin{equation}
e_{50} = \mathrm{median}(e_i),
\end{equation}
\begin{equation}
\zeta = \frac{\mathrm{median}(\sigma_i)}{\mathrm{mean}(\sigma_i)},
\end{equation}
and
\begin{equation}
\sigma_i = \sqrt{\sigma_G^2+e_i^2-e_{50}^2},
\end{equation}
where the errors on individual stellar radial velocities are denoted as $e_i$. Uncertainties on this estimator are calculated from bootstrapping \citep{Efron:1979}. For the defined sample of stars in NGC 1333, this estimate of $\sigma_0$ is 1.04 $\pm$ 0.18 km s$^{-1}$. This broader confidence interval is indicative of the relative instability of this estimator for small samples \citep{astroMLText}.

Finally, based on the previous estimates of the intrinsic velocity dispersion, we trim all stars with radial velocity uncertainty $>$ 0.5 km~s$^{-1}$. This allows us to approximate the errors as roughly uniform (and small, compared to the intrinsic velocity dispersion) and simply calculate

\begin{equation}
\sigma_0 = \sqrt{\sigma_{r}^2 - e_{50}^2},
\end{equation}
where $\sigma_{r}$ is some robust estimator of the dispersion. Using either the Median Absolute Deviation \citep[MAD][]{Muller:2000} or $\sigma_G$ for $ \sigma_{r} $ produces comparable results with $\sigma_0$ =1.12 $\pm$ 0.18 km s$^{-1}$. This estimate for the dispersion is actually slightly greater than the estimate just from the sample standard deviation of the untrimmed data, as many of the stars with radial velocity uncertainty between 0.5 and 1.1 km~s$^{-1}$ lie near average velocity. The uncertainty on this estimate is again fairly large, because we have significantly reduced the amount of data used in this estimate. Note that \citet{Cottaar:2014c} also trim all stars with radial velocity uncertainty $>$ 0.5 km~s$^{-1}$ in their analysis of IC 348.

The estimates of the intrinsic velocity dispersion of the stars in NGC 1333 from \textsc{velbin}, the outlier-resistant analytic estimate, and the velocity dispersion of the error-trimmed subset are therefore all consistent with one another, within their relatively large uncertainties. We proceed with the \textsc{velbin} estimate of $\sigma_0$ =  0.92$\pm$0.12~km~s$^{-1}$, as this estimate uses the most information.

\subsection{Comparison with Low-Density Gas}

 \begin{figure*}
\includegraphics[width=0.99\textwidth]{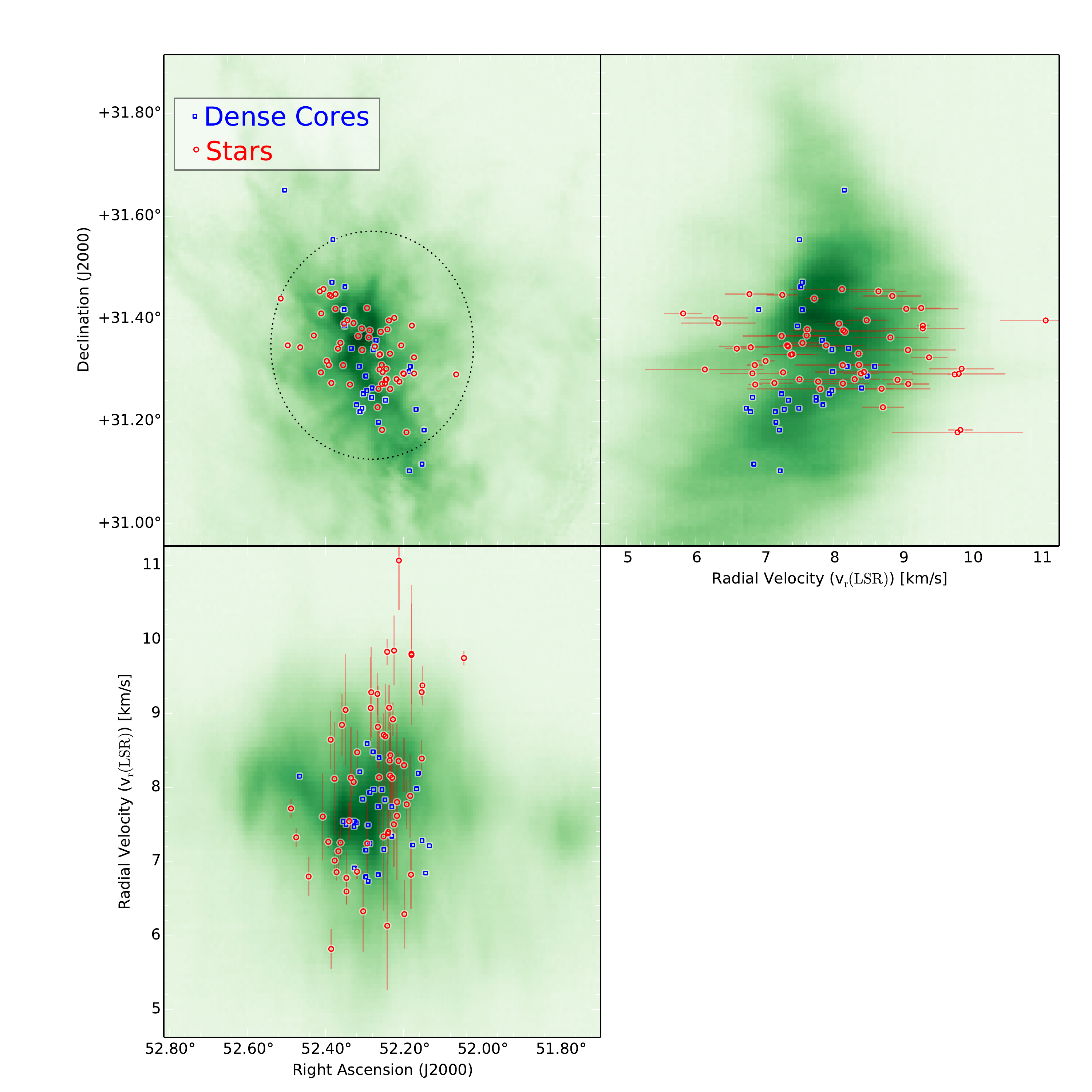}
\caption{Positions and radial velocities of IN-SYNC stars (red) and \nthp\ cores \citep[blue; from][]{Kirk:2007} in NGC 1333. Shown in the green colorscale is the intensity of $^{13}$CO (1-0) gas integrated over velocity (top-left), Right Ascension (top-right) and Declination (bottom-left). Error bars on the IN-SYNC stars show the 1$\sigma$ uncertainty on radial velocity for these stars; the velocity uncertainty on the \nthp\ cores is much smaller (typically $<$ 0.05 km s$^{-1}$) and these errors are suppressed for clarity. The stellar population is contained within a radius of 800\arcsec (0.97 pc), which is shown in the dotted circle (top-left).}
\label{fig:diffusegas}
\end{figure*}

\autoref{fig:diffusegas} shows the positions and velocities of the dense cores and IN-SYNC stars compared with the low-density gas tracer, $^{13}$CO (1-0), which shows the general cloud gas. The cores and stars have similar, although not identical, spatial distributions. The cores have a small velocity dispersion (the one-dimensional velocity dispersion is 0.51~$\pm$~0.05~km~s$^{-1}$) and are strongly correlated with the highest intensity regions of the diffuse gas; this confirms what other studies have found -- dense cores are not moving ballistically with respect to their surrounding diffuse gas \citep{Walsh:2004, Kirk:2007, Kirk:2010}.

The radial velocity errors on the stars would tend to diminish the appearance of any real correlation between the stellar velocities and that of the diffuse gas. Nonetheless, the structure of stars with low radial velocity errors reveals some cases where the stars are not well correlated with the diffuse gas. Comparing the radial velocity of the IN-SYNC stars with the first moment (i.e., the intensity-weighted mean) of the emission profile from the cloud gas (both $^{13}$CO (1-0) and C$^{18}$O (3-2)) along the line of sight toward each of the stars shows no correlation between the diffuse gas velocity and the stellar velocity. This lack of a correlation arises because the first moment of the cloud gas is essentially the same at all locations in the cluster, while the stars have a broad spread in radial velocities. 

\begin{figure}
\includegraphics[width=0.47\textwidth]{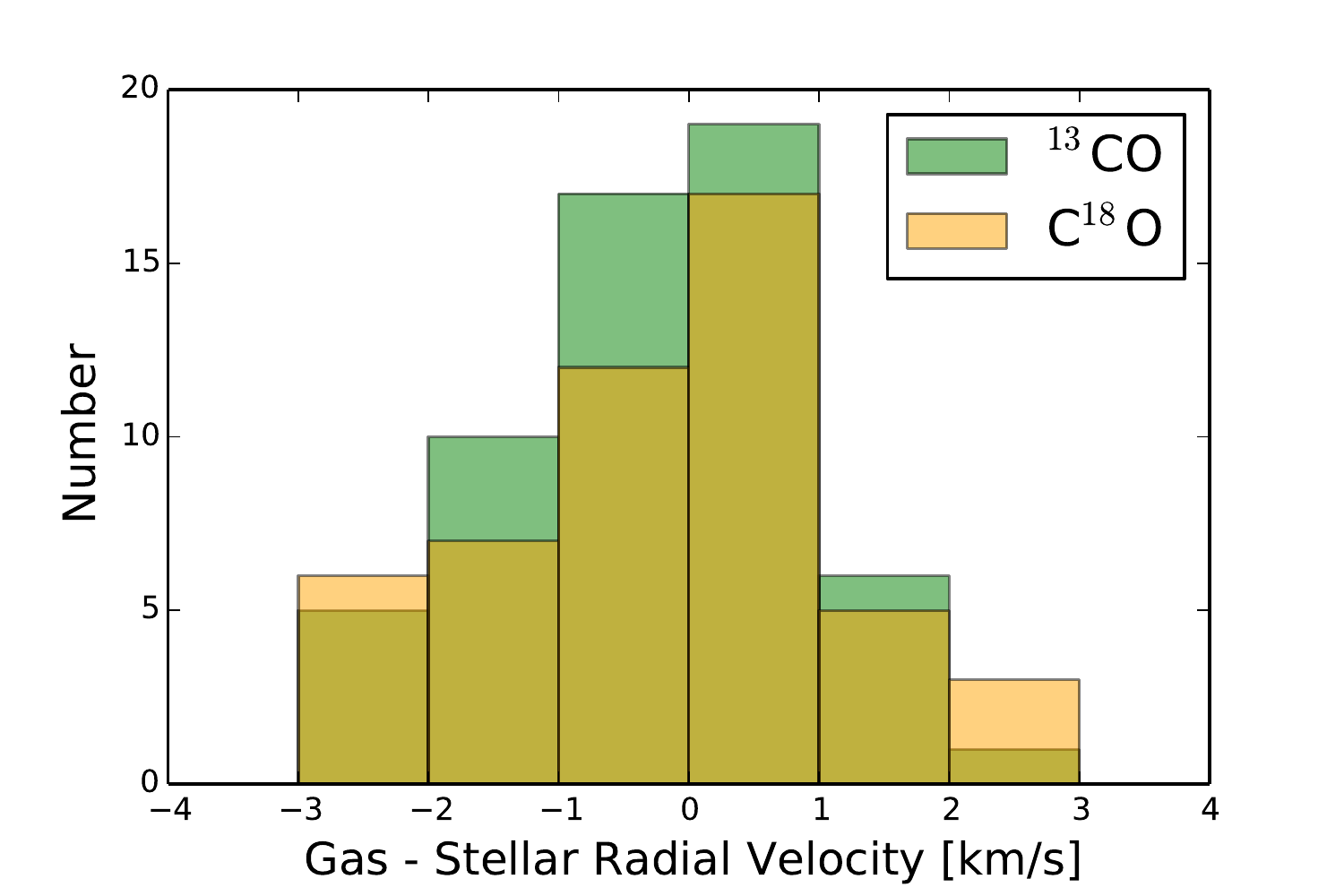}
\caption{The velocity difference between each star's radial velocity and the centroid velocity of $^{13}$CO (1-0) (green) and C$^{18}$O (3-2) (orange) at that position. Five stars fall outside the region covered by the C$^{18}$O (3-2) map and are not shown.} 
\label{fig:velocitydiff}
\end{figure}

\autoref{fig:velocitydiff} displays the difference between stellar radial velocity and the centroid (1st moment) velocity of the $^{13}$CO (1-0) and C$^{18}$O (3-2) gas. In both cases, the mean offset is close to zero, and the standard deviation of the offsets are roughly 1 km s$^{-1}$, which turns out to be comparable with the line-width of the $^{13}$CO (1-0) gas.

\subsection{Comparison of Stellar and Dense Core Velocity Dispersions}

\begin{figure}
\includegraphics[width=0.47\textwidth]{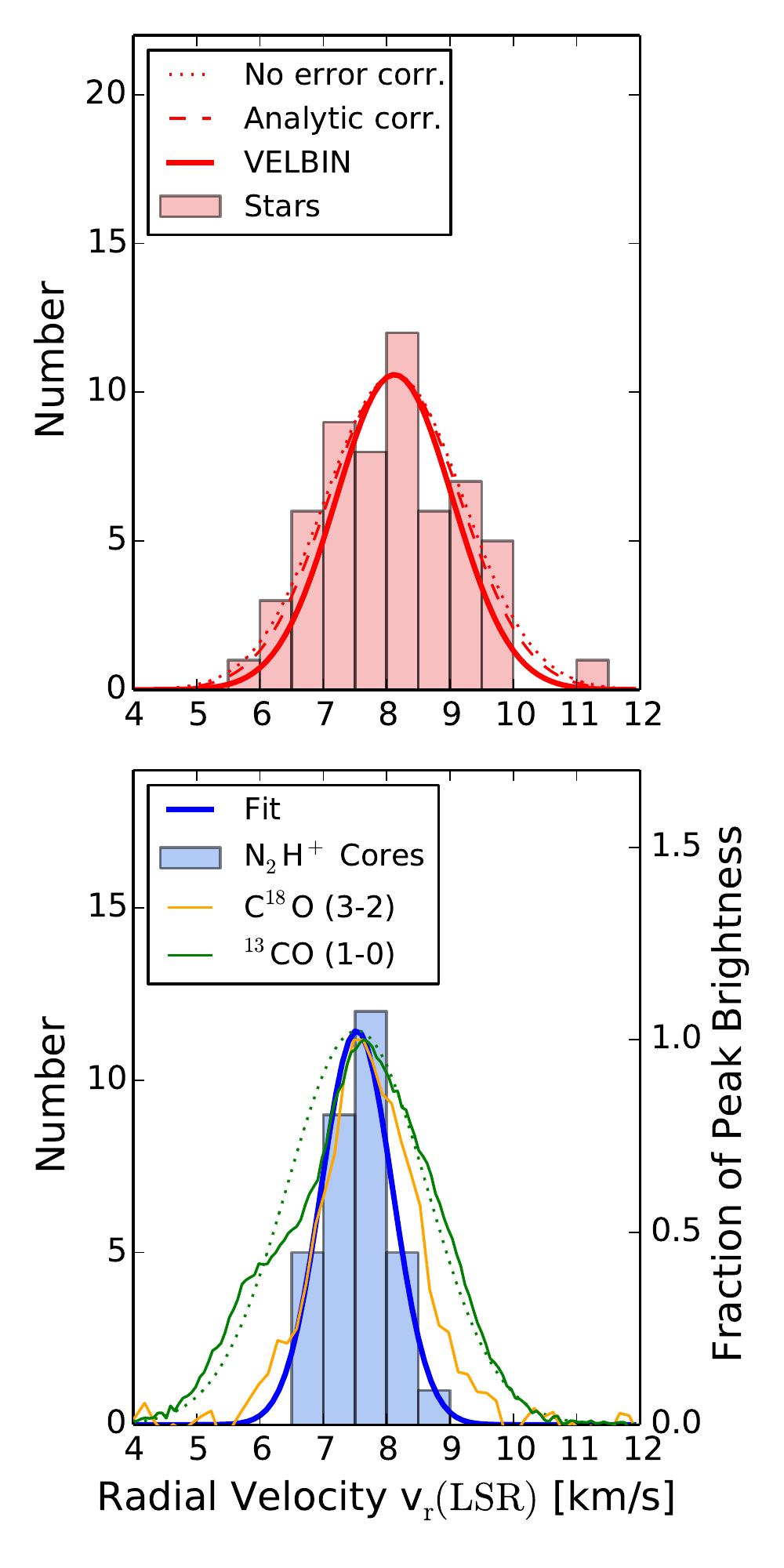}
\caption{Comparison between the radial velocity distributions for \nthp\ cores in NGC 1333 \citep[bottom; from][]{Kirk:2007} and IN-SYNC stars (top). The stellar histogram is broadened by the errors on the radial velocity determination (which are negligible for the \nthp\ cores). Lines show the inferred width without account for errors (dotted), \textsc{velbin} (solid) and the analytical correction for non-uniform errors given in the text (dashed). For comparison with the core velocity dispersion, the average spectrum of $^{13}$CO (1-0) (green; low-density tracer) and C$^{18}$O (orange; high-density tracer) are over-plotted on the \nthp\ core velocity distribution, scaled to the amplitude of the core velocity dispersion distribution. The dotted (green) line shows the line width of $^{13}$CO (1-0) with $\sigma$ = 1.1 km  s$^{-1}$ as estimated in \autoref{sec:virial}.}
\label{fig:velocitycomparison}
\end{figure}

This measurement of the velocity dispersion in NGC 1333 represents the earliest measurement of this quantity\footnote{Arguably the stars in the ONC studied by \citet{Tobin:2009} are of comparable age, but due to its higher mass, the dynamical age (i.e., age/dynamical time) of the ONC is significantly greater than the dynamical age of NGC 1333.}. NGC 1333 is still actively forming stars, with dense gas cores that are either starless or host to early (Class 0/I) protostars. These cores lie in roughly the same spatial area as the young stars measured by IN-SYNC but have a much tighter velocity dispersion. Taking only the cores from \citet{Kirk:2007} that lie within our defined spatial boundary, the one-dimensional velocity dispersion is 0.51~$\pm$~0.05~km~s$^{-1}$.

Note that the determination of the dense core velocity dispersion is not corrected for the effects of binarity. One might imagine, for instance, that a binary system with separate accretion envelopes could produce a skewed line profile in \nthp. In the case of two envelopes with high relative velocities, the \nthp\ line profile could separate into two velocity components, although this situation is indistinguishable from multiple widely-separated cores within the beam along the line of sight. 

We can directly compare the velocity dispersion of the population of dense \nthp\ cores and the population of stars observed with IN-SYNC. Because the two populations lie in similar positions within the cluster, this comparison is relating the velocity dispersion of stars just before they form (presumably stars inherit the systemic velocity of the dense cores out of which they form) and about 1-2 Myr after their birth. 

\autoref{fig:velocitycomparison} shows the histograms of velocities for dense cores and stars. The distributions have consistent central velocities within their uncertainties ($v_{\mathrm{cen}}(\mathrm{cores})$ = 7.59$\pm$0.09~km~s$^{-1}$, $v_{\mathrm{cen}}(\mathrm{stars})$ = 8.02$\pm$0.31~km~s$^{-1}$). Because the stellar distribution appears broader due to the significant and non-uniform uncertainties on the radial velocity determinations, we plot the Gaussian distributions inferred in \autoref{sec:stellar-dispersion} as well as the simple fit to the velocity distribution. The dense core velocity dispersion is significantly less broad than the stellar velocity dispersion inferred with \textsc{velbin}. The dense cores have a similar velocity dispersion to the line-width seen in the dense gas tracer (C$^{18}$O (3-2)), but significantly smaller dispersion than the diffuse gas (C$^{13}$O (1-0)).

\subsection{Virial State of Stars and Cores}
\label{sec:virial}

The full virial equation for a molecular cloud is

\begin{equation}
\frac{1}{2}\ddot{I}  = 2(\mathcal{T} - \mathcal{T_S}) + \mathcal{M} + \mathcal{W},
\label{eqn:fullvirial}
\end{equation}
where $\ddot{I}$ denotes the acceleration of the expansion/contraction of the cloud, $\mathcal{T}$ is the kinetic energy of particles and gas within the cloud, $\mathcal{T_S}$ is the surface term (surface pressure), $\mathcal{M}$ is the energy in the magnetic field and $\mathcal{W}$ is the gravitational potential energy. A simple virial analysis normally assumes that one can neglect the surface term and the magnetic field, and that a cloud's expansion or contraction is not accelerating ($\ddot{I} = 0$) so that
\begin{equation}
- \mathcal{W} = 2\mathcal{T}.
\end{equation}
Solving this equation for a spherical distribution with a power-law density distribution, $\rho(r) \propto r^{-k}$, gives \citep[see][]{Bertoldi:1992} a virial velocity dispersion, $\sigma_{\mathrm {vir}}$, via
\begin{equation}
\sigma_{\mathrm {vir}} = \sqrt{\frac{a M G}{5 R}}
\label{eq:simplevirial}
\end{equation}
where $R$ is the radius, $M$ the mass of the region under consideration, and $a$ is a geometric factor of order unity:
\begin{equation}
a = \frac{1-k/3}{1-2k/5}.
\end{equation}

Evaluating the virial state NGC 1333 therefore requires knowledge of the mass of the cluster at a given radius. This mass includes both the gas mass and the stellar mass. The projected gas distribution (\autoref{fig:coldensity}) shows significant sub-structure and non-azimuthal symmetry, and so estimates of the virial velocity assuming a smooth distribution will necessarily be only a rough approximation. 

\begin{figure}
\includegraphics[width=0.49\textwidth]{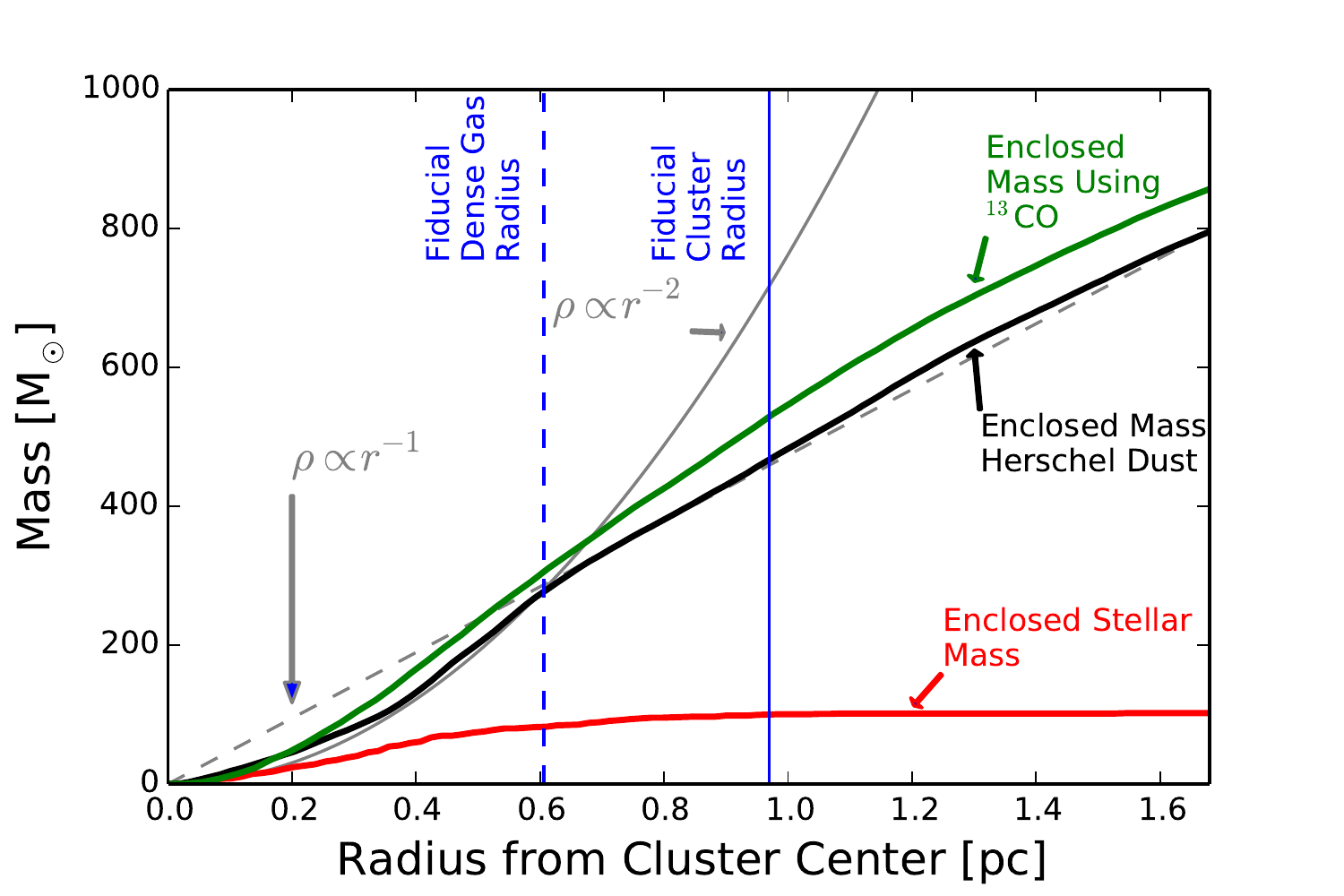}
\caption{Mass enclosed within a given radius of the center of NGC 1333. Two different estimates are compared, that from CO (green) and that from the \emph{Herschel} dust map (black). The enclosed mass profiles assuming a density power law $\rho \propto r^{-1}$ (constant column density) or $\rho \propto r^{-2}$ (normalized to the dust map at 1.65 pc and 0.6 pc, respectively) are shown in gray; the observed surface density profile is intermediate between these values. Also shown is the contribution from stars assuming that every star in the input IN-SYNC catalog has a mass of 0.5 M$_{\odot}$ (red line). Vertical lines denote radii of interest.} 
\label{fig:massenclosed}
\end{figure}

We compare two estimates of the gas mass within NGC 1333. The first is the mass estimate derived from the \emph{Herschel} column density map calculated in \autoref{sec:dustcolumn}. The second is using the COMPLETE $^{12}$CO and $^{13}$CO maps to derive the excitation temperature of the CO and then use the X-factors calculated for this region by \citet{Pineda:2008} to convert the CO intensity to a total mass. The comparison between these two methods is shown in \autoref{fig:massenclosed}. The conversion between CO and total mass depends on relations which were not calibrated in the central region of NGC 1333, since the extinction map used in the conversion was unreliable there \citep{Pineda:2008}. For this reason, we use the mass estimated from the \emph{Herschel} dust column density map, although the difference between the two profiles is not large.

\autoref{fig:massenclosed} also shows the enclosed mass for $\rho(r) \propto r^{-1}$ and $\rho(r) \propto r^{-2}$ profiles. The observed mass profile is intermediate between these two simple profiles, matching the $k=1$ profile at radii $>$ 0.6 pc, and close to the $k=2$ profile at radii $<$ 0.6 pc. This change in profile occurs at the edge of the region of high column density (i.e. $\sim$ 15 mags of A$_{V}$) seen in the \emph{Herschel} column density map (see \autoref{fig:coldensity}). The majority of the stellar population is contained with 800\arcsec (0.97 pc at a distance of 250 pc), and we use this as the fiducial cluster radius. 

IN-SYNC does not provide a complete census of all the stars in NGC 1333, so an estimate of the total stellar mass from these data requires significant extrapolation. Our input catalog contains 205 objects in the region considered. Low-mass regions such as this one have a typical mean stellar mass of 0.5 \Msun. In order to show how this mass is distributed in \autoref{fig:massenclosed}, we simply assign 0.5 \Msun\ to every star in the IN-SYNC catalog, giving a total mass of 102 \Msun. For comparison, \citet{Lada:1996} estimate a total stellar cluster mass of  45 \Msun\ over a similar region to that which is considered here. This is obviously a fairly rough approximation, but the stellar mass is much smaller than the gas mass, and therefore has relatively little influence on the virial velocity of the cluster.

\begin{figure}
\includegraphics[width=0.49\textwidth]{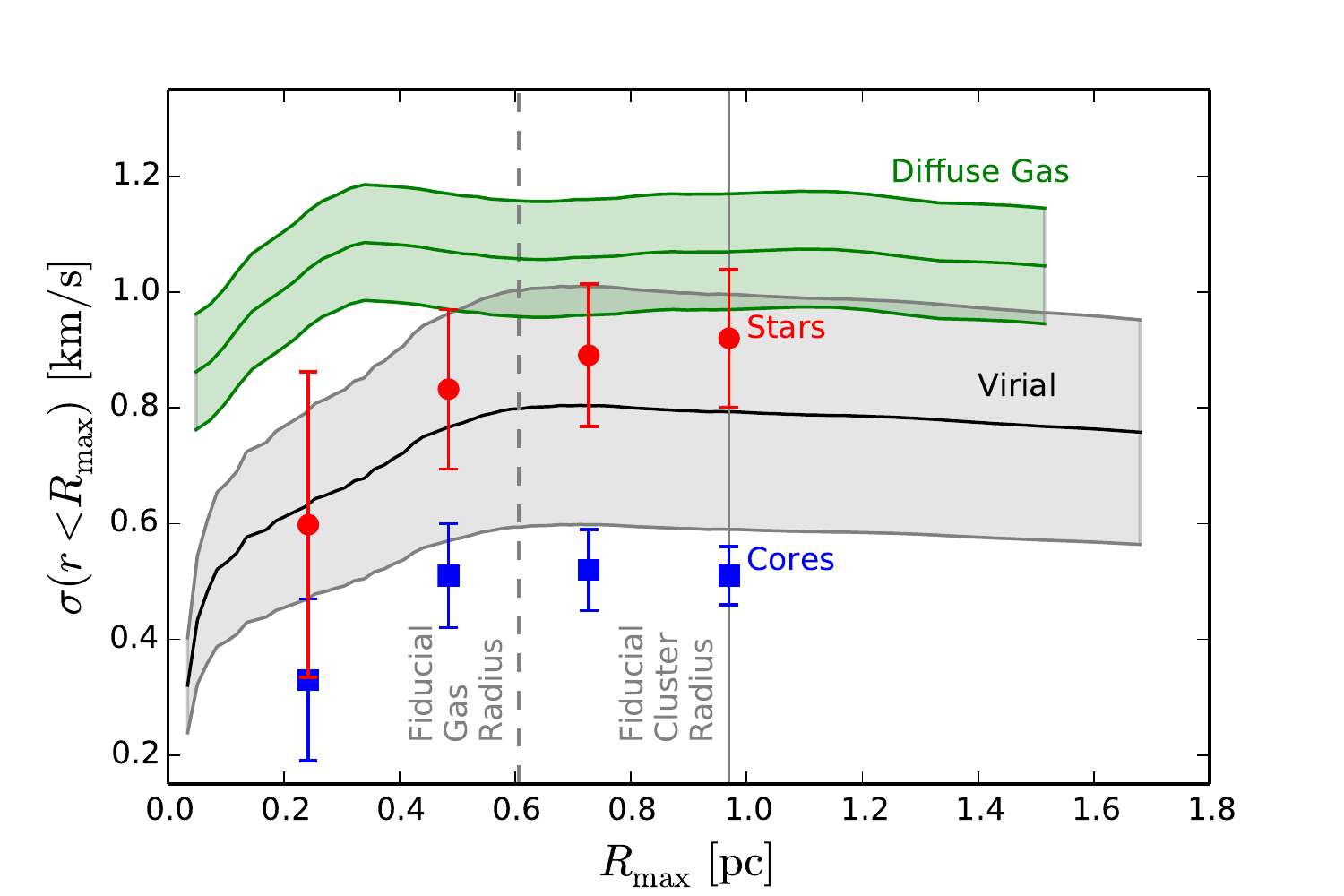}
\caption{Velocity dispersion of the stars (red circles), dense cores (blue squares), diffuse gas ($^{13}$CO (1-0); green), and the expected velocity dispersion (black) if the cluster were in virial equilibrium (ignoring magnetic fields and external pressure). The velocity dispersions are shown as a function of $R_{\mathrm{max}}$, and are calculated by considering all objects/positions interior to $R_{\mathrm{max}}$. Vertical lines denote radii of interest. Shaded regions and error bars show estimates of the 1$\sigma$ uncertainty of these values.} 
\label{fig:velocitydisp}
\end{figure}

The virial velocity given by \autoref{eq:simplevirial} can be evaluated as a function of radius. Since the projected mass profile implies that the true density profile is between $\rho(r) \propto r^{-2}$ and $\rho(r) \propto r^{-1}$, we use a single value of $\rho(r) \propto r^{-1.5}$ to provide a continuous value for the virial velocity. This leads to a value of $a$ = 1.25. We show the result of this calculation in gray in \autoref{fig:velocitydisp}, along with an estimate of the uncertainty, which is dominated by the systematic uncertainty on the mass estimate, which is a combination of an uncertain zero-point in our column density map, the uncertain dust emissivity, and our assumption of a single temperature component along the line of sight. We adopt a factor of two uncertainty on the mass to account for these sources of uncertainty. The virial velocity takes on a roughly constant value beyond 0.6 pc (as would be expected for the $\rho(r) \propto r^{-1}$ profile which is observed at large radii) of $\sigma_{\mathrm{vir}} = 0.79 \pm 0.20$~km~s$^{-1}$.

Another estimate of the virial velocity comes from assuming that the diffuse gas is roughly in virial equilibrium, and therefore that the virial velocity dispersion of NGC 1333 can be estimated from the line width of the diffuse gas. As shown in \autoref{fig:velocitycomparison}, the line profile for  $^{13}$CO (1-0) integrated over the cluster region is non-Gaussian. We adopt the common approach of measuring the FWHM of the emission and then converting to an (effective) $\sigma$ for a Gaussian distribution. The result of measurement as a function of radius is shown in green in \autoref{fig:velocitydisp}. There are several important caveats with this measurement: the presence of significant outflow energy and momentum in NGC 1333 \citep[e.g.][]{Lefloch:1998a,Knee:2000,Plunkett:2013} means that the CO in NGC 1333 may be super-virial; on the other hand, $^{13}$CO (1-0) may not be optically thin, so optical depth effects may increase the observed line width. 

Finally, we can estimate the intrinsic velocity dispersion for both stars and dense cores, binned within a given radius. For this purpose we use \textsc{velbin} for the stars and the sample standard deviation for the cores (as we did in \autoref{sec:stellar-dispersion} and \autoref{sec:stellar-dispersion} for the full cluster). These measurements are displayed in \autoref{fig:velocitydisp}.

\autoref{fig:velocitydisp} shows that the various estimates of the velocity dispersion approach constant values beyond about 0.4 pc. Inside this radius there is a suggestive dip in all measurements of the velocity dispersion, although for any individual tracer this is not a statistically significant decrease. At small radii, neglecting the surface term, $\mathcal{T_S}$, in \autoref{eqn:fullvirial} is obviously incorrect since the surrounding mass at of the cluster gas makes a significant contribution. Furthermore, it is difficult to interpret the measured radial velocity of stars at radii smaller than the orbital radii. Therefore we take the velocity dispersions measured at large radii as indicative of the true state of the cluster.

The stellar velocity dispersion of 0.92 $\pm$ 0.12 km s$^{-1}$ is therefore consistent with our estimate of the cluster's virial velocity (0.79 $\pm$ 0.20 km s$^{-1}$) and the line width of the surrounding diffuse gas (1.1 $\pm$ 0.1 km s$^{-1}$); the dense cores have a velocity dispersion of 0.51 $\pm$ 0.05 km s$^{-1}$, less than the velocity dispersion of the diffuse gas, and consistent with sub-virial.

\section{Discussion}

\subsection{Sub-clusters in NGC 1333}
\label{sec:subclusters}
\citet{Lada:1996} first noted the bimodal spatial distribution of young stars in NGC 1333, which exhibits distinct northern and southern cluster in the stars observable in the near-infrared. The relatively unbiased survey of \citet{Gutermuth:2008} confirmed the existence of these two sub-clusters, which can also be seen in \hyperref[fig:hrd]{Figure~\ref*{fig:hrd}a} as the clusters of stars around Declination 31$\arcdeg$.37 and 31$\arcdeg$.27, respectively. This clustering is less prominent in \autoref{fig:diffusegas} since the northern cluster contains fainter stars on average; fainter stars tend to have larger radial velocity uncertainties and are therefore suppressed in this figure.

Given the presence of sub-clusters, does it make sense to consider the velocity dispersion and virial velocity of the cluster as a whole? Using the \citet{Gutermuth:2008} division of NGC 1333 into two clusters separated by a Declination of 31$\arcdeg$.3, we use \textsc{velbin} to calculate the $v_{c}$ and $\sigma_0$ for the northern and the southern clusters. For the northern cluster, $v_{c}$ = 7.9 $\pm$ 0.42 km s$^{-1}$ and  $\sigma_0$ = 0.92 $\pm$ 0.18 km s$^{-1}$. For the southern cluster, $v_{c}$ = 8.1 $\pm$ 0.40 km s$^{-1}$ and  $\sigma_0$ = 0.94 $\pm$ 0.15 km s$^{-1}$. There is thus no detectable kinematic difference between the two sub-clusters, and considering them separately produces roughly the same result for the stellar velocity dispersion. We therefore proceed with considering the velocity dispersion of the full cluster.

\subsection{Velocity Gradients in NGC 1333}
\citet{Quillen:2005} report a velocity gradient in $^{13}$CO (1-0) of 1 km s$^{-1}$ in the north-south direction across NGC 1333. However, this gradient is fit in a larger region than we consider here, extending 15\arcmin\ further south, and the magnitude of the gradient is strongly influenced by the substantial blue-shift in emission in the southern portion of their field. The diffuse gas shown in \autoref{fig:diffusegas} certainly displays some large-scale structure, but does not appear to exhibit a single consistent gradient. This is confirmed with higher-excitation CO lines; \citet{Bieging:2014} present CO (2-1) and $^{13}$CO (2-1) maps of NGC 1333 and speculate that the patchy variations in centroid velocity seen in these tracers could be ascribed to a cloud-cloud collision.

We find no statistically significant velocity gradient in the stellar velocities across NGC 1333. For instance, considering just a north-south gradient, the best-fit velocity gradient across the region is -2.9 $\pm$ 2.2 km s$^{-1}$ degree$^{-1}$ (12.6 $\pm$ 9.6 km s$^{-1}$ pc$^{-1}$) . Since the area under consideration is 1/3 of a degree, the uncertainty on the magnitude of this gradient is on order of 0.7 km s$^{-1}$, leaving open the possibility of a gradient which is large enough to contribute to the observed spread in radial velocities seen in the region. Ultimately, we rely on the fact that we cover the same spatial region with our three tracers (diffuse gas, stars, and dense cores) and assume that any large-scale gradients or patchy structure effect the three tracers to a similar degree.

\subsection{Comparison with Simulations}

\citet{Proszkow:2009} show that clusters which start with a sub-virial velocity distribution become slightly super-virial as the cluster collapses, while an initially virial velocity distribution remains roughly constant. Non-spherical elongation of the cluster, particularly along the line of sight, makes it difficult to distinguish between the case of sub-virial velocity dispersion and projection effects. Nonetheless, our results are broadly consistent with these simulations which start sub-virial, and then quickly become virial.

\citet{Offner:2009} examined the velocity dispersion of young stars in a simulation of turbulent star formation. They find that star clusters forming in turbulent virialized clouds naturally begin with sub-virial velocity dispersions, however these sub-virial velocity dispersions persist for one free-fall time. The dynamical time, $t_{\mathrm{dyn}}$ = R/$\sigma_{\mathrm {vir}}$, of NGC 1333 is 1.1 Myr, so our result is in conflict with this prediction. Furthermore, the stars in the \citet{Offner:2009} simulation retain a strong correlation with the centroid velocity of the gas in which they are embedded; our stars show no such correlation. This simulation seems to predict the behavior of the starless and protostellar cores traced by N$_2$H$^{+}$, but not the behavior of the 1-2 Myr old stars that are observable in the near-infrared.

Both \citet{Kruijssen:2012} and \citet{Girichidis:2012} have recently studied the dynamical state of young stars in simulations. \citet{Kruijssen:2012}, in simulations with rather more mass than NGC 1333 (10$^4$ \Msun\ versus the 10$^3$ \Msun\ in NGC 1333), find that after one free-fall time most of the stars are in sub-clusters with relatively little gas left, and the stars are in viral equilibrium. Our study suggests that gas expulsion does not play a critical role in the virialization of young stars; the mass of NGC 1333 is still dominated by gas, and yet its young stars are virialized. \citet{Girichidis:2012} study a slightly earlier stage, when only 20\% of the mass is in protostars. In this study, only the central portions of the cluster are virialized, the external stars contribute to a sub-viral velocity distribution for the cluster as a whole. We see no evidence of this trend; the central regions and the outskirts of the cluster appear to have a similar viral state.

\subsection{Explaining the Velocity Dispersion Difference}

The difference between the 0.92~km~s$^{-1}$  stellar velocity dispersion and the 0.5~km~s$^{-1}$ dense core velocity dispersion can be explained in a number of ways. First, consider the additional terms in the virial equation that influence the gas and stars differently. After their formation, stars will cease to feel both the external pressure and the magnetic field. Since the dynamical time of NGC 1333 is 1.1 Myr, it is possible for the velocity dispersion of 1-2 Myr old stars to evolve subsequent to their formation. 

We first consider the influence external pressure on the region. We can calculate the external pressure using the formula from \citet{Lada:2008}:
\begin{equation}
P_{\mathrm{S}} = 4.5 \times 10^3 \phi_G k A^2_V,
\end{equation}
where $\phi_G$ is a geometric factor of order 1 and $A_V$ is the average column density in the region outside the cloud, which we can estimate as 2 mags of $A_V$ for NGC 1333 from the COMPLETE extinction map. The energy due to surface pressure is simply
\begin{equation}
\mathcal{T_S}  = 4 \pi R^3 P_S.
\end{equation}
Our estimate for $\mathcal{T_S}$ is therefore $1.3\times10^{45}$ erg, which is less than 10\% of the potential energy in the cluster ($|\mathcal{W}| = 3.3\times10^{46}$ erg). The external pressure does not seem to be significant.

We cannot estimate the strength of the magnetic field in NGC 1333 directly, but if we assume that the difference between the dense core kinetic energy and the energy of the potential well in which they are embedded comes purely from the magnetic field, then we can calculate the strength of that field as 

\begin{equation}
\mathcal{M} = \frac{B^2 R^3}{6},
\end{equation}
which gives B = 66$\mu$G. This is a reasonable strength for the magnetic field in a region as dense as NGC 1333. \citet{Crutcher:2012} define an empirical relation between the strength of the magnetic field and the density of a region. For regions denser than the threshold particle density of $n_0$ = 300 cm$^{-3}$,

\begin{equation}
B = B_0 \left(\frac{n}{n_0}\right)^{0.65}
\end{equation}
where $B_0$ = 10$\mu$G. Assuming spherical symmetry, the particle density, $n$, of NGC 1333 is roughly $6\times10^3\ $cm$^{-3}$, and thus $B$ = 70$\mu$G. 

Whether a magnetic field of this strength would actually provide support will depend at least partly on the field's morphology. In general, an ordered magnetic field will not provide support against collapse along magnetic field lines, but any realistic magnetic field configuration will involve turbulence translating support against collapse into all directions \citep[see][and references therein]{McKee:2007}. The exact details depend on how tangled the magnetic field is, and how much energy density is in large-scale field components; further investigation with simulations is required.

The diffuse gas has a broader velocity dispersion, and it might be expected to be more strongly influenced by the magnetic field. However, the diffuse gas could well be out of virial equilibrium due to the injection of energy from outflows from the existing young stars in NGC 1333.

It is therefore quite possible that NGC 1333 has a magnetic field strong enough to explain the dense cores' sub-virial motion. In this picture, the dense cores' velocities are constrained by the magnetic field; upon their formation the stars are freed from the magnetic field and evolve to the velocity dispersion dictated by gravity. 

The most difficult problem with this explanation is that the young stellar population in NGC 1333 does not seem to be in the relaxed, centrally-condensed configuration expected for a virialized stellar population. As discussed in \autoref{sec:subclusters}, the stellar population exhibits significant structure in the form of two sub-clusters. Furthermore, an analysis of the radial density profile of the young stars in NGC 1333 by \citet{Gutermuth:2008} finds a flat central surface density profile, which is interpreted as evidence of very little dynamical evolution.

We consider a second explanation: that the cluster NGC 1333 might be in a state of global collapse. The IN-SYNC stars, which are older than the current dense cores, therefore formed when the cluster was more extended than it is today, presumably with an initially sub-virial dispersion (like that possessed by the current dense cores). As the cluster globally collapsed, the potential energy of the stellar configuration was transformed into additional kinetic energy, rendering the stars dynamically hotter than the current population of dense cores. \citet{Cottaar:2014c} have recently proposed that IC 348 is in a state of global collapse based on an independent line of reasoning. This explanation would therefore lend support to the idea that young clusters are generally in a state of global collapse.

In this picture the sub-virial velocity dispersion of dense cores can be explained if dense cores form at the convergent point of large-sale turbulent flows \citep{Elmegreen:2007, Gong:2011} or if the cores form from a small number of velocity-coherent structures \citep[perhaps filaments;][]{Hacar:2011}. The dense cores in NGC 1333 do exhibit coherence in position-velocity space (see top-right panel of \autoref{fig:diffusegas}), so this is a reasonable explanation. If the IN-SYNC stars come from similarly sub-structured initial conditions, where those structures are now undergoing global collapse, then perhaps the stellar velocity dispersion could be inflated within the sub-structures before the sub-structure is erased. In this model it is unclear why the stellar sub-clusters in NGC 1333 have consistent central velocities; the na\"{i}ve expectation in the case of global collapse is that they would have different bulk motions. Additional modeling of this point is required. 

\section{Conclusions}

With the aid of the first high-resolution multi-object near-infrared spectrograph, APOGEE, we have measured, for the first time, the velocity dispersion of stars in an embedded young low-mass cluster where the stellar velocity dispersion can be directly compared with that of the dense star-forming cores and the diffuse gas in which both are embedded. Our results for NGC 1333 show that the 1-2 Myr old (Class II, where classification is possible) stars have an intrinsic velocity dispersion of 0.92 $\pm$ 0.12 km s$^{-1}$ after correcting for measurement uncertainties and the influence of binaries. The velocity dispersion of these young stars is significantly greater than the velocity dispersion of the dense cores (0.51 $\pm$ 0.05 km s$^{-1}$) in the same region.

Unlike for the dense cores, the stars studied here are moving ballistically with respect to the low density gas; there is no correlation between the two on a point-to-point comparison, although the mean velocity of the population is the same as the mean velocity of the diffuse gas and the width of the velocity distributions are similar. The stellar velocity dispersion is roughly virial, considering just the gravitational potential produced by the stars and gas in NGC 1333; in comparison the dense core velocity dispersion is sub-virial. Two possible (though not mutually exclusive or exhaustive) explanations for these results are (1) the presence of a magnetic field with strength of order 70$\mu$G having a strong influence on the velocity dispersion of the dense cores, or (2) a globally collapsing cluster with initial sub-structure. In both these scenarios, the velocity dispersion of the stars must increase quickly. Star-formation theories and simulations should strive to reproduce similar velocity dispersions for young stars roughly 1-2 Myr ($\sim$ 1-2 $t_{\mathrm{dyn}}$) after their birth.

\section{Acknowledgements}

JBF performed the analysis contained herein and wrote the manuscript. MC developed and ran the spectral analysis routines producing the stellar parameters used in this paper. KRC, JCT and MRM conceived the programÕs scientific motivation and scope, led the initial ancillary science proposal, oversaw the projectÕs progress and contributed to the analysis of the stellar parameters; KRC also led the target selection and sample design process, and provided assistance with the analysis and interpretation of the APOGEE spectra. HAG provided access to the dense gas data and help with interpreting the comparison between stellar and gas velocities. DLN assisted in the interpretation of the APOGEE data products and reduction algorithms, particularly those related to radial velocity measurements. NDR provided comparison with the gas and stellar velocities in Orion. KMF assisted with target selection. KGS contributed to the discussion of velocity spreads. SDC and GZ oversaw the design of the APOGEE plates utilized for IN-SYNC observations. SRM, MFS, and PMF contributed to defining the scope and implementation plan for this project, and with JCW  developed and provided high level leadership for the broader APOGEE infrastructure and survey that enabled this science. LR contributed to the development of the input catalog for NGC 1333.

This research made use of Astropy, a community-developed core Python package for astronomy \citep{astropy:2013}. This research has made use of NASA's Astrophysics Data System. We thank Stella Offner, Alyssa Goodman, Alvaro Hacar, Ralf Klessen and Mark Heyer for insightful discussions. We thank our referees for insightful reports which improved this paper.

Funding for SDSS-III has been provided by the Alfred P. Sloan Foundation, the Participating Institutions, the National Science Foundation, and the U.S. Department of Energy Office of Science. The SDSS-III web site is \url{http://www.sdss3.org/}.

SDSS-III is managed by the Astrophysical Research Consortium for the Participating Institutions of the SDSS-III Collaboration including the University of Arizona, the Brazilian Participation Group, Brookhaven National Laboratory, Carnegie Mellon University, University of Florida, the French Participation Group, the German Participation Group, Harvard University, the Instituto de Astrofisica de Canarias, the Michigan State/Notre Dame/JINA Participation Group, Johns Hopkins University, Lawrence Berkeley National Laboratory, Max Planck Institute for Astrophysics, Max Planck Institute for Extraterrestrial Physics, New Mexico State University, New York University, Ohio State University, Pennsylvania State University, University of Portsmouth, Princeton University, the Spanish Participation Group, University of Tokyo, University of Utah, Vanderbilt University, University of Virginia, University of Washington, and Yale University.

\bibliographystyle{apj}
\bibliography{/Users/jonathanfoster/Dropbox/Bibdesk/Bibdesk}

\clearpage

\LongTables
\capstartfalse

\begin{deluxetable*}{rrrrrrr}
\tabletypesize{\tiny}
%\rotate
\tablewidth{0pt}
\tablecaption{Candidate young stars in NGC 1333 \label{table:alltarget}}
\tablehead{\colhead{ObjID}     & \colhead{2MASSID}                                         & \colhead{RA}   &
	           \colhead{Dec}                & \colhead{J}                    & \colhead{H}   &                    
	           \colhead{K}    \\
		\colhead{}                 &\colhead{}                               &  \colhead{[deg.]}    
		               &  \colhead{[deg.]}         &  \colhead{[mag]}   &  \colhead{[mag]}			& \colhead{[mag]}\\
	       }
\startdata

1	&	2M03274053+3115392	&	51.9189	&	31.2609	&	  8.88	&	  8.68	&	  8.60\\
2	&	2M03281101+3117292	&	52.0459	&	31.2915	&	 12.44	&	 11.46	&	 11.03\\
3	&	2M03281518+3117238	&	52.0633	&	31.2900	&	 16.56	&	 15.71	&	 15.19\\
4	&	2M03282206+3110429	&	52.0919	&	31.1786	&	 13.87	&	 12.61	&	 11.86\\
5	&	2M03282839+3116273	&	52.1183	&	31.2743	&	 14.62	&	 13.93	&	 13.62\\
6	&	2M03283107+3117040	&	52.1295	&	31.2845	&	 16.82	&	 15.35	&	 14.08\\
7	&	2M03283258+3111040	&	52.1357	&	31.1848	&	 17.41	&	 15.43	&	 14.49\\
8	&	2M03283641+3128569	&	52.1517	&	31.4825	&	 15.24	&	 13.83	&	 13.18\\
\enddata
\tablecomments{\autoref{table:alltarget} will be published electronically.}
\end{deluxetable*}

\capstarttrue

\clearpage

\capstartfalse

\begin{deluxetable*}{rrrrrrrrrrrrrr}
\tabletypesize{\tiny}
%\rotate
\tablewidth{0pt}
\tablecaption{Fit parameters for all observed APOGEE Spectra in NGC 1333 \label{table:allspectra}}
\tablehead{\colhead{ObjID}                               & \colhead{MJD}                               & \colhead{$v_{r}$}   &
	           \colhead{$\sigma$}                & \colhead{$v$ sin $i$}                    & \colhead{$\sigma$}   &                    
	           \colhead{$T_{\mathrm{eff}}$}         & \colhead{$\sigma$}             & \colhead{log $g$}                &
	            \colhead{$\sigma$}         &  \colhead{$r_{H}$} 			  &  \colhead{$\sigma$}   &
	            \colhead{red-$\chi^{2}$}               & \colhead{S/N}                         	           \\
		\colhead{}                                            & \colhead{[days]}	                     &  \multicolumn{2}{c}{[km s$^{-1}$]}    
		               &  \multicolumn{2}{c}{[km s$^{-1}$]}         &  \multicolumn{2}{c}{[K]}                    &  \multicolumn{2}{c}{[log$_{10}$ cm s$^{-2}$]}  & \colhead{}				& \colhead{}				&
	        \colhead{}					& \colhead{}					\\
	       }
\startdata

1	&	56671.05549	&	-29.24	&	  0.95	&	 90.48	&	  1.38	&	6110.00	&	 40.00	&	  3.66	&	  0.05	&	  0.24	&	  0.02	&	5.4	&	190\\
1	&	56674.05859	&	-25.11	&	  0.87	&	 99.16	&	  1.50	&	5850.00	&	 70.00	&	  3.68	&	  0.08	&	  0.06	&	  0.01	&	9.5	&	220\\
1	&	56563.49856	&	-29.28	&	  1.20	&	 90.96	&	  1.78	&	6120.00	&	 50.00	&	  3.74	&	  0.05	&	  0.17	&	  0.02	&	3.7	&	160\\
1	&	56607.32096	&	177.85	&	  0.02	&	  0.01	&	  0.58	&	3890.00	&	 30.00	&	  4.15	&	  0.09	&	  0.00	&	  0.00	&	551.5	&	180\\
1	&	56561.31439	&	-28.21	&	  0.67	&	 91.93	&	  1.07	&	6120.00	&	 30.00	&	  3.70	&	  0.03	&	  0.21	&	  0.01	&	9.3	&	280\\
1	&	56236.22732	&	-32.07	&	  0.71	&	 88.04	&	  0.98	&	5980.00	&	 20.00	&	  3.60	&	  0.02	&	  0.22	&	  0.01	&	28.4	&	270\\
1	&	56315.22693	&	-33.26	&	  0.65	&	 91.20	&	  0.88	&	6150.00	&	 40.00	&	  3.85	&	  0.05	&	  0.18	&	  0.01	&	35.5	&	320\\
2	&	56671.05549	&	  9.91	&	  0.22	&	  3.01	&	  2.26	&	3540.00	&	 20.00	&	  3.93	&	  0.08	&	  0.13	&	  0.04	&	5.3	&	 50\\
\enddata
\tablecomments{\autoref{table:allspectra} will be published electronically. Fit parameters are the effective temperature ($T_{\mathrm{eff}}$), the veiling ($r_H$), the rotational velocity ($v \sin{i}$), the surface gravity ($\log(g)$) and the radial velocity ($v_r$) of the star.}
\end{deluxetable*}

\capstarttrue

\clearpage

\capstartfalse

\begin{deluxetable*}{rrrrrrrrrrrrrrr}
\tabletypesize{\tiny}
%\rotate
\tablewidth{0pt}
\tablecaption{Average parameters for all stars in NGC 1333 \label{table:allstar}}
\tablehead{\colhead{ObjID}                                           & \colhead{$v_{r}$}   &
	           \colhead{$\sigma$}                & \colhead{$v$ sin $i$}                    & \colhead{$\sigma$}   &                    
	           \colhead{$T_{\mathrm{eff}}$}         & \colhead{$\sigma$}             & \colhead{log $g$}                &
	            \colhead{$\sigma$}         &  \colhead{$r_{H}$} 			  &  \colhead{$\sigma$}   &
                \colhead{absJ} & \colhead{nobs} &
	            \colhead{ngood\tablenotemark{a}}               & \colhead{var\tablenotemark{b}}     \\
		\colhead{}                                              &  \multicolumn{2}{c}{[km s$^{-1}$]}    
		               &  \multicolumn{2}{c}{[km s$^{-1}$]}         &  \multicolumn{2}{c}{[K]}                    &  \multicolumn{2}{c}{[log$_{10}$ cm s$^{-2}$]}  & \colhead{}				& \colhead{}				&
	       \colhead{[mag]} &
           \colhead{}			& \colhead{}		& \colhead{}					\\
	       }
\startdata

1	&	-27.80	&	  0.43	&	 92.94	&	  0.68	&	 6100	&	   20	&	  3.69	&	  0.02	&	  0.15	&	  0.01	&	  2.10	&	 7	&	 4	&	0\\
2	&	  9.75	&	  0.10	&	  6.23	&	  0.60	&	 3540	&	   10	&	  3.98	&	  0.03	&	  0.11	&	  0.01	&	  4.50	&	 7	&	 6	&	0\\
4	&	 -9.85	&	  2.79	&	 81.74	&	  4.13	&	 6570	&	   30	&	  3.38	&	  0.07	&	  0.33	&	  0.05	&	  3.60	&	 7	&	 6	&	0\\
5	&	   \nodata	&	   \nodata	&	   \nodata	&	   \nodata	&	  \nodata	&	  \nodata	&	   \nodata	&	   \nodata	&	   \nodata	&	   \nodata	&	   \nodata	&	 3	&	 0	&	0\\
8	&	   \nodata	&	   \nodata	&	   \nodata	&	   \nodata	&	  \nodata	&	  \nodata	&	   \nodata	&	   \nodata	&	   \nodata	&	   \nodata	&	   \nodata	&	 3	&	 0	&	0\\
10	&	  9.38	&	  0.27	&	 27.81	&	  0.45	&	 3320	&	   10	&	  3.90	&	  0.04	&	  0.17	&	  0.02	&	  5.60	&	 7	&	 7	&	0\\
11	&	  8.39	&	  0.26	&	 53.68	&	  0.40	&	 3760	&	   10	&	  3.64	&	  0.02	&	  0.12	&	  0.01	&	  3.20	&	 7	&	 6	&	0\\
12	&	  9.29	&	  0.09	&	 11.94	&	  0.26	&	 3900	&	   10	&	  4.21	&	  0.02	&	  0.13	&	  0.01	&	  4.00	&	 7	&	 7	&	0\\
\enddata
\tablenotetext{a}{Number of spectra meeting the quality criteria given in the text.}
\tablenotetext{b}{Binary flag indicating if the source has a significantly variable radial velocity as defined in the text (1 for variable, 0 for non-variable).}
\tablecomments{\autoref{table:allstar} will be published electronically. Fit parameters are the effective temperature ($T_{\mathrm{eff}}$), the veiling ($r_H$), the rotational velocity ($v \sin{i}$), the surface gravity ($\log(g)$) and the radial velocity ($v_r$) of the star.}
\end{deluxetable*}

\capstarttrue

\end{document}